# Temperature and Carbon Assimilation Regulate the Chlorosome Biogenesis in Green Sulfur Bacteria

Joseph Kuo-Hsiang Tang[1,*], Semion K. Saikin[2], Sai Venkatesh Pingali[3], Miriam M. Enriquez[4], Joonsuk Huh[2], Harry A. Frank[4], Volker S. Urban[3], Alán Aspuru-Guzik[2]

[1]School of Chemistry and Biochemistry, Clark University, Worcester, MA 01610 USA, [2]Department of Chemistry and Chemical Biology, Harvard University, Cambridge, MA 02138 USA, [3]Center for Structural Molecular Biology, Biology and Soft Matter Division, Oak Ridge National Laboratory, Oak Ridge, TN 37831 USA, [4]Department of Chemistry, University of Connecticut, Storrs, CT 06269 USA

Running title: Metabolic regulation of chlorosome

*To whom correspondence should be addressed: Tel: 1-614-316-7886, Fax: 1-508-793-8861, E-mail: jtang@clarku.edu

**Abstract**

Green photosynthetic bacteria adjust the structure and functionality of the chlorosome – the light absorbing antenna complex – in response to environmental stress factors. The chlorosome is a natural self-assembled aggregate of bacteriochlorophyll (BChl) molecules. In this study we report the regulation of the biogenesis of the *Chlorobaculum tepidum* chlorosome by carbon assimilation in conjunction with temperature changes. Our studies indicate that the carbon source and thermal stress culture of *Cba. tepidum* grows slower and incorporates less BChl *c* in the chlorosome. Compared with the chlorosome from other cultural conditions we investigated, the chlorosome from the carbon source and thermal stress culture displays: (a) smaller cross-sectional radius and overall size; (b) simplified BChl *c* homologues with smaller side chains; (c) blue-shifted $Q_y$ absorption maxima and (d) a sigmoid-shaped circular dichroism (CD) spectra. Using a theoretical model we analyze how the observed spectral modifications can be associated with structural changes of BChl aggregates inside the chlorosome. Our report suggests a mechanism of metabolic regulation for chlorosome biogenesis.

Abbreviation

BChl, Bacteriochlorophyll; CD, circular dichroism; FMO protein, Fenna-Matthews-Olson protein; SANS, small-angle neutron scattering; TCA cycle, tricarboxylic acid cycle

## INTRODUCTION

Green photosynthetic bacteria, which include green sulfur bacteria and green non-sulfur bacteria, can live in low-light or extremely low-light environments (1-3). In order to efficiently harvest light in such conditions green photosynthetic bacteria use the chlorosome, a natural optical antenna complex. The solar energy absorbed by the chlorosome is converted to molecular excitations and then transferred to the reaction center, where a transmembrane electron potential is generated. The chlorosome contains hundreds of thousands of bacteriochlorophylls (BChls) (4-6) that are self-assembled into supramolecular aggregates with few proteins associated. Several reports have implied the great potential of the chlorosome for it use in biohybrid and biomimetic devices (7-9).

Biogenesis of the chlorosome from green sulfur bacteria has been reported to be regulated by multiple environmental factors including light intensities (10), temperature (11, 12), light intensities combined with temperatures (13) and sulfite concentrations (13, 14). In this work we study how the chlorosome in the green sulfur bacterium *Chlorobaculum* [Cba.] *tepidum* is adapted to the type of carbon source and thermal stress conditions. Carbon assimilation and electron transport is known to be a bottle-neck of photosynthesis (15). In order to store solar energy, green sulfur bacteria use the reductive (reverse) tricarboxylic acid (TCA) cycle for carbon fixation (2, 16, 17). In contrast to the oxidative (forward) TCA cycle, which is energetically favorable, the reductive TCA cycle uses ATP and NAD(P)H produced via photosynthetic electron transport for assimilating inorganic carbon and synthesizing acetyl-CoA. Green sulfur bacteria are known to grow mixotrophically with acetate, pyruvate and a few organic carbon sources (2). Studies indicate that mixotrophic-grown green sulfur bacteria operate mainly the reductive TCA cycle for assimilating pyruvate, because of low carbon flux from pyruvate to acetyl-CoA, and operate both oxidative and reductive TCA cycles for assimilating acetate (18) (**Fig. S1**). Thus green sulfur bacteria exhibit better growth with acetate whereas require more energy to assimilate pyruvate.

In this study we analyzed the structural and morphological modifications of the *Cba. tepidum* chlorosome upon changes in temperatures and nutrition; viz. organic carbon sources, using an array of techniques such as small-angle neutron scattering (SANS), UV-visible absorption, circular dichroism (CD), 77 K fluorescence emission spectra, and mass spectra. To obtain more structural insights on BChl packing in the chlorosome from the different cultures we computed absorption and CD spectra of BChl aggregates composing the chlorosome.

## MATERIALS AND METHODS

*Materials*. All solvents and reagents were obtained from standard commercial sources and used as received. The thermophilic green sulfur bacterium *Cba. tepidum* was cultured anaerobically and phototrophically using the medium as reported (2, 18). The cultures were grown in low-intensity light ($20 \pm 0.2$ μmole/$m^2$/s) inoculated with 1% cell cultures grown at 50 $^oC$ or at 30 $^oC$ in the late exponential growth phase. Cell cultures were monitored using optical density (OD) of a cell suspension at 625 nm, where the absorbance of photosynthetic pigments is minimal (18). The cell growth rate was estimated from the exponential phase of cell growth. Chlorosomes were purified as reported (7, 19). Unless otherwise mentioned, chlorosomes in this study were prepared in 20 mM Tris-HCl buffer at pH 7.8, and chlorosome samples were incubated with dithiothreitol (DTT). Bacteriochlorophyll *c* was prepared by extraction from chlorosomes using methanol followed by C18 Sep-Pak cartridge (Waters) treatment (12). All measurements for chlorosomes were performed in darkness or under extremely low light intensity. Data reported in this paper represented the mean of at least 5 independent measurements for chlorosomes isolated from cultures grown under similar conditions.

*77 K Fluorescence Emission Spectra*. Fluorescence emission measurements were carried out at 77 K using a Jobin-Yvon Horiba Fluorolog-3 model FL3-22 equipped with a Hamamatsu R928P detector and a 450 W ozone-free Osram XBO xenon arc lamp. The fluorimeter was set to right-angle detection relative to the excitation light. The samples were adjusted to have an OD between 0.10 and 0.35 in a 4 mm cuvette at the excitation wavelength of 735 nm and contained 70 % glycerol (v/v). The samples were held in plastic cuvettes that were 1 cm (excitation path) by 4 mm (emission path) and attached to a custom-made holder that was lowered slowly into an optical immersion dewar cryostat (Kontes Custom Shop) containing liquid nitrogen. The emission spectra were recorded using slit widths on the excitation and emission monochromators corresponding to a bandpass of 5 nm. The emission signal was corrected for the instrument response function using a correction file generated by a reference photodiode.

*Room Temperature Spectral Measurements.* The UV-visible absorption spectra reported here were recorded using a Shimadzu UV-1800 spectrophotometer. The room temperature fluorescence emission spectra were recorded at 25 $^{o}$C with various excitation wavelengths using a FluoroMax-3 Spectrofluorimeter (J.Y. Horiba, Ltd.). The fluorescence emission spectra of the samples were measured at a 90$^{o}$ angle relative to the incident light to minimize the transmitted light, and chlorosomes with $OD_{748}$ = 0.08 were measured to minimize the inner filter effect. The CD spectra for the chlorosome in a 1.5 mm path quartz cuvette were recorded between 350 and 850 nm at 25 $^{o}$C using a Jasco J-810 CD spectrometer. LC-MS and LC-MS/MS analyses were performed on an Agilent 6520 Q-TOF LC/MS/MS mass spectrometer system with an ESI source (12). The hydrodynamic diameter of the chlorosome was measured with a ZetaSizer Nano ZS (Malvern Instruments Inc., UK) using dynamic light scattering (DLS) (7, 12).

*Small-angle Neutron Scattering*. Small-angle neutron scattering (SANS) measurements were made using the Bio-SANS instrument at the High Flux Isotope Reactor, Oak Ridge National Laboratory (Oak Ridge, TN) and data were analyzed using approaches reported previously (7, 20, 21). Scattering intensity profiles were modeled using Igor Pro routines made available by the NIST Center for Neutron Research (22). SANS data of the chlorosome were analyzed using modified Guinier analysis for a rod-like particle (7, 20) using the conventional relation

$$q \times I(q) = I(0)\exp\left(\frac{-q^2 R_c^{\,2}}{2}\right);\ R = \sqrt{2} \times R_c, \qquad \textbf{(1)}$$

where $q$ is scattering vector or wave-vector transfer, $R_c$ is the cross-sectional radius of gyration and $R$ is the radius of a rod-like particle of infinite length.

The scattering data of the chlorosome were also analyzed with a bi-axial ellipsoid of rotation fit (22). A scale factor, which is proportional to square of (the particle volume) term and particle concentration (i.e., number of such particles), was included in the data fit.

## RESULTS

*Grow Cba. tepidum at Various Temperatures and with Different Carbon Sources*

We first investigated *Cba. tepidum* grown in 12 different cultural conditions with varied temperatures (constant temperature, temperature up-shifted, and temperature down-shifted) and carbon sources ($HCO_3^-$, $HCO_3^-$/acetate and $HCO_3^-$/pyruvate) (**Table S1**). The culture growth rates, $Q_y$ absorption maxima, room temperature fluorescence emission maxima, and hydrodynamic diameters of the chlorosome are summarized in **Table S2**. The results indicate that the growth rate for cultures supplied with various carbon sources follows the trend: $HCO_3^-$/acetate cultures > $HCO_3^-$ cultures > $HCO_3^-$/pyruvate cultures. Moreover, the overall size of the chlorosome from $HCO_3^-$/pyruvate cultures is smaller than the chlorosome from $HCO_3^-$/acetate cultures. Culture 6 (the temperature down-shifted culture grown on pyruvate/$HCO_3^-$), which in our understanding experiences maximal environmental stress, in addition to slowest growth rate and smallest overall size, exhibited a substantial blue-shift of the $Q_y$ absorption maximum and the room temperature fluorescence peak. In the rest of the paper we compare the structural, optical and functional properties of the chlorosome from culture 6 with chlorosomes from other cultures and also with chlorosomes from *Cfl. aurantiacus*.

A. *Structural Information*

To acquire more structural insights on the chlorosome from culture 6 and other cultures grown on various carbon sources and temperatures, we employed liquid chromatography/mass spectrometry (LC/MS) to analyze the BChl *c* homologues in the chlorosome and small-angle neutron scattering (SANS) to investigate the morphology of the chlorosome.

(I) *LC/MS Measurements*

We analyzed the ratio of BChl *c* homologues in the chlorosome via LC/MS. **Fig. 1** shows the data from 6 of 12 cultures. **Fig. 1*A*** shows the HPLC chromatogram for BChl *c* homologues in the chlorosome with detection wavelength at 640 nm, and **Fig. 1*B*** reports the percentage of BChl *c* homologues in the chlorosome analyzed by mass peak area of LC/MS. As reported in the literature (12, 23, 24), major BChl *c* homologues in the *Cba. tepidum* chlorosome are [8-ethyl, 12-methyl]-farnesyl-BChl *c* (*m/z* 793.5), [8-ethyl, 12-ethyl]-farnesyl-BChl *c* (*m/z* 807.5), [8-propyl, 12-ethyl]-farnesyl-BChl *c* (*m/z* 823.5), [8-isobutyl, 12-ethyl]-farnesyl-BChl *c* (*m/z* 839.5) and [8-neopentyl, 12-ethyl]-farnesyl-BChl *c* (*m/z* 853.5). **Fig. 1*A*** shows the co-elution of [8-ethyl, 12-ethyl]-farnesyl-BChl *c*/[8-propyl, 12-ethyl]-farnesyl-BChl *c* (Rt 5.35 min), [8-propyl, 12-ethyl]-farnesyl-BChl *c*/[8-isobutyl, 12-ethyl]-farnesyl-BChl *c* (Rt 5.77 min), and [8-isobutyl, 12-ethyl]-farnesyl-BChl *c*/[8-neopentyl, 12-ethyl]-farnesyl-BChl *c* (Rt 6.30 min). Also, **Fig. 1*A*** indicates that culture 6 exhibits simpler BChl *c* homolog distributions with very small peak in the retention time at 5.8 min (corresponding to BChl *c* homologues with larger substituents at C-8) (12), and **Fig. 1*B*** shows that more than 70% of BChl *c* homologues in the chlorosome from culture 6 are [8-ethyl, 12-ethyl]-farnesyl-BChl *c* (*m/z* 807.5), in contrast to more heterogeneous distribution of BChl *c* homologues in other cultures. Culture 6 contains simpler BChl *c* homologues, which is akin to the distribution of BChl *c* homologues in *Cfl. aurantiacus*, which only has [8-ethyl, 12-methyl]-BChl *c* due to the absence of genes encoding C-$8^2$, C-$12^1$ methyltransferases (25).

(II) *SANS Measurements*

We employed SANS for probing the morphology (i.e., size and shape) of the chlorosome. Our previous studies highlighted the power of SANS in elucidating structural features critical to the function of chlorosomes (7, 20, 21, 26). **Fig. 2*A*** shows scattering data of the chlorosome from cultures 1, 3, 4, 6, 7 and 9, with the data merged from three experimental configurations covering the *q*-range 0.001–0.70 $Å^{-1}$. All chlorosome samples for SANS measurements were prepared in 100% $D_2O$ buffer (20 mM Tris-HCl at pH 7.8). **Fig. 2*B*** shows that chlorosomes from all cultures exhibit a rod-like shape, indicating that the morphology of the chlorosome from culture 6 is similar to the chlorosome from other cultures. **Table 1** lists the values of the parameters of modified Guinier analysis for a rod-like particle (Eq. (**1**) in Materials and Methods) (7, 20) performed on the chlorosome from all cultures. The cross-sectional radius of gyration of the chlorosome from culture 6 is significantly smaller as compared to other cultures. The results follow the trend observed in the dynamic light scattering measurements (**Table S2**).

**Table 1. Scattering parameters of the chlorosome determined with modified Guinier analysis**

| Chlorosomes | ***I*(0)** | **$R_c$ (Å)[1]** | ***R* (Å)[2]** | **$R_c$*$Q_{max}$** |
|---|---|---|---|---|
| Chlorosome 1 | 0.22 ± 0.04 | 126 ± 4 | 178 ± 6 | 0.92 ± 0.05 |
| Chlorosome 3 | 0.38 ± 0.03 | 123 ± 3 | 174 ± 4 | 0.91 ± 0.04 |
| Chlorosome 4 | 0.78 ± 0.03 | 127 ± 3 | 180 ± 4 | 0.93 ± 0.04 |
| Chlorosome 6 | 0.14 ± 0.02 | 92 ± 4 | 130 ± 6 | 0.92 ± 0.03 |
| Chlorosome 7 | 0.15 ± 0.03 | 122 ± 3 | 173 ± 4 | 0.90 ± 0.03 |
| Chlorosome 9 | 0.51 ± 0.04 | 126 ± 4 | 178 ± 6 | 0.92 ± 0.04 |

[1]$R_c$ is the cross-sectional radius of gyration for a particle.
[2]***R*** is the radius of a rod-like particle of infinite length defined as: $R = \sqrt{2}R_c$

**Fig. 2*C*** shows the scattering data analyzed with a bi-axial ellipsoid of rotation fit (22). The cross-sectional radius of the chlorosome analyzed with modified Guinier model (*R* value in **Table 1**) agrees with the bi-axial ellipsoid of rotation fit (perpendicular semi-axis value in **Table 2**), further indicating that the chlorosome from culture 6 appears to have smaller cross-sectional radius (i.e., more flat) compared with the chlorosome from other cultures.

**Table 2. Scattering parameters of the chlorosome determined with bi-axial ellipsoid of rotation fit, estimated volume of the chlorosome and estimated number of BChl *c* in the chlorosome**

| Chlorosomes | Perpendicular semi-axis (nm) | Parallel semi-axis (nm) | Volume (x $10^3$ $nm^3$) | Number of BChl *c* (x $10^3$) |
|---|---|---|---|---|
| Chlorosome 1 | 19.2 ± 0.4 | 221.1 ± 41.2 | 344.2 ± 77.9 | 163 ± 40 |
| Chlorosome 3 | 18.8 ± 0.3 | 184.1 ± 32.4 | 274.2 ± 56.7 | 129 ± 27 |
| Chlorosome 4 | 18.9 ± 0.5 | 241.4 ± 24.5 | 363.4 ± 55.8 | 173 ± 28 |
| Chlorosome 6 | 15.4 ± 0.5 | 162.8 ± 30.0 | 163.8 ± 40.3 | 77 ± 21 |
| Chlorosome 7 | 18.9 ± 0.6 | 182.4 ± 25.0 | 275.6 ± 54.8 | 132 ± 27 |
| Chlorosome 9 | 18.7 ± 0.4 | 171.0 ± 22.0 | 252.0 ± 43.0 | 119 ± 20 |

Taken LC/MS and SANS data together, the chlorosome from culture 6 exhibits simpler BChl *c* homologues and smaller cross-sectional radius compared with the chlorosome from other cultures.

B. *Optical Properties*

(I) *Absorbance and CD Measurements*

The structural studies described above indicate a different morphology of the chlorosome from culture 6 compared with the chlorosome from other cultures. To understand further the changes in the internal structure of the chlorosome from the stressed culture, we compared the chlorosome from two temperature down-shifted cultures: culture 4 (grown on $HCO_3^-$/acetate) and culture 6 (grown on $HCO_3^-$/pyruvate). **Fig. 3*A*** shows the UV-visible spectra of culture 4 and culture 6, where culture 6 displays a blue-shifted $Q_y$ maximum and smaller chlorosome peak intensity. The chlorosome from culture 6 displays smaller intensity of the $Q_y$ absorption band, **Fig. 3*B*,** implying that less BChls are contained in the chlorosome. Additionally, we observed that the $Q_y$ maxima of the culture 6 chlorosome and the *Cfl. aurantiacus* chlorosome were almost identical, **Fig. 3*C***, with the culture 6 chlorosome exhibiting a broader peak. The latter fact is likely due to various BChl *c* homologues present in the culture 6 chlorosome (27).

**Fig. 4*A*** shows CD of the chlorosome from culture 4 and culture 6 with different spectral features. Three types of CD spectra have been introduced previously (28): type I (+/–) (positive at the shorter wavelength to negative at the longer wavelength), type II (–/+) (negative at the shorter wavelength to positive at the longer wavelength) and mixed-type (–/+/–) CD spectra (coexistence of type I and type II). CD spectra of the culture 6 chlorosome and the culture 4 chlorosome belong to the type II (–/+) and mixed-type (–/+/–) CD spectra, respectively. The chlorosome from other cultures investigated in this report, except culture 6, also exhibited a mixed-type CD spectrum (data not shown). Mixed-type CD spectra are also reported for most of the studies of the *Cba. tepidum* chlorosome (29). We noticed that the culture 6 chlorosome and the *Cfl. aurantiacus* chlorosome exhibited similar type II CD features, **Fig. 4*B***, with a slightly blue-shifted on the negative band of the culture 6 chlorosome. Both type I (+/–) and type II (–/+) CD spectra have been reported in the *Cfl. aurantiacus* chlorosome (7, 21, 30-32) and the *Cba. tepidum* (mutant) chlorosome (33, 34). It has been suggested that BChl *c* aggregates fold in the opposite direction (35), which gives either a type I or type II CD spectra, and that the presence of both types of BChl *c* aggregates in the chlorosome lead to a mixed-type CD spectra.

(II) *77 K Fluorescence Emission Spectra*

To characterize the stress-associated changes in the excitation energy transfer we compared low temperature, $T$ = 77 K, fluorescence emission spectra of the whole cell and the chlorosome from culture 4 and culture 6. For the chlorosome spectra the rate of excitation transfer from BChl *c* in the chlorosome (780 – 784 nm) to BChl *a* in the baseplate complex can be estimated with the ratio of the peak area (the baseplate complex/chlorosome). In the whole cell spectra this ratio (the baseplate complex or the FMO protein/chlorosome) can also be affected by energy transfer from the baseplate to the FMO protein and reaction center.

Green sulfur bacteria are known to exhibit redox-dependent energy transfer (1, 2). Thus, cultures and chlorosomes was incubated in 100 mM freshly prepared DTT in the anaerobic chamber overnight as described previously (12). **Fig. 5*A*** shows a stronger relative emission from the baseplate energy domain of the stressed chlorosome, with peak ratio 25% more compared to the chlorosome from culture 4 (**Fig. S2**). This can be an indication of an enhanced chlorosome-to-baseplate energy transfer in culture 6, despite the blue-shift of the chlorosome absorption peak. The emission spectra from the whole cells show that exciton population of the baseplate in stressed culture is smaller, **Fig. 5*B***, with peak ratio 50% less compared to the chlorosome from culture 4 (**Fig. S3**). We hypothesize that the temperature/nutrition stress induces structural changes in the light-harvesting complex beyond the chlorosome, which may enhance the energy funneling from the baseplate. However, the detailed analysis of this process goes beyond the scope of the current study.

**Model**

BChl aggregates in the chlorosome are organized as a mixture of tubular and lamellar shapes (36). Recent cryo-electron microscopy studies suggest that inside the chlorosome tubular aggregates (rolls) are composed in larger concentric assemblies (35, 37). **Fig. 6** illustrates our schematic understanding of aggregate packing in an idealized chlorosome, where $L_{1\text{-}3}$ are three semi-axes of the ellipsoidal body and $R$ is a radius of a tubular aggregate. Therefore, the chlorosome length $L_1$ determines the maximal length of rolls in the structure, $L_3$ limits the maximal radius of a roll, and the ratio $L_2 / L_3$ controls the contribution of lamella vs. roll shapes.

Among several existing models for BChl aggregation in chlorosomes (35, 38-41), we utilized the one introduced by Ganapathy and coworkers (35, 41). In addition to conventional nuclear magnetic resonance (NMR) and cryo-electron microscopy studies, this structure has also been supported using 2-dimensional polarization fluorescence microscopy (42). We have shown recently that the timescales characterizing exciton dynamics in this structures agree with the ones obtained in time-resolved measurements of the chlorosome (43, 44). In this model the BChl stacks form a two-dimensional lattice shown in **Fig. 7*A***. Subsequently, the lattice is folded into a roll. We considered two types of folding: structure I – BChl stacks form concentric rings, **Fig. 7*B***, and structure II – BChl stacks are parallel to the symmetry axis of the roll, **Fig. 7*C***. While the former structure was predicted for a mutant chlorosome that consists of BChl *d* pigments (35), the same group of researchers suggested that the latter structure describes a wild type chlorosome (41). The latter structure also qualitatively agrees with the model proposed previously (38) and used for the description of CD spectra of chlorosomes (45). For the purpose of simplicity in the following discussion, we neglect the distance variation between *syn-anti* dimers, which can introduce a shift of the absorption band but cannot result in the change of the type of the CD spectra.

From the bi-axial ellipsoid of rotation fit for SANS measurements and the chlorosome model shown in **Figs. 6-7** we estimate the volume of the chlorosome and the amount of BChl *c* in different cultures and report the results in **Table 2**. Herein we assume that the spacing between BChl aggregates is equal to 2.1 nm (35) and that approximately a half of the chlorosome volume is occupied by BChls (1). The number of BChl *c* homologues in the chlorosome has also been estimated by thin section and freeze-fracture electron microscopy (2,400 – 12,000 BChls in the *Cfl. aurantiacus* chlorosome) (4), atomic force microscopy (215,000 ± 80,000 BChls in the *Cba. tepidum* chlorosome) (46), fluorescence correlation spectroscopy (~140,000 BChls in the *Cba. tepidum* chlorosome and ~96,000 BChls in the *Cfl. aurantiacus* chlorosome) (5) and confocal fluorescence microscopy (50,000 – 100,000 in the *Cba. tepidum* chlorosome) (6). Though our estimations are qualitative, the values we obtain are comparable to other reported results. The number of BChls in the stressed culture 6 is approximately a half of the number in other cultures (**Table 2**).

The shift of the $Q_y$ absorption maximum of the chlorosome can originate from the difference in electronic excitation spectra of single BChl *c* homologues and also modifications of BChl assemblies on different length scales. In contrast, single molecule contribution to CD spectra of the chlorosome is very small and most of the signal is due to the supramolecular assembly.

*Absorption and CD Spectra of Molecular Aggregates*

We utilized single BChl rolls as a representative model for the absorption and CD spectra of the chlorosome, assuming that the absorption signal is dominated by the rolls with the largest number of molecules. The latter assumption is relaxed for CD spectra because the signal is stronger for rolls with a larger curvature. Thus, we limited the maximal size of representative rolls to be approximately a half of the size of a chlorosome obtained from the SANS measurements, **Table 1**.

In order to predict possible changes in the BChl aggregation in the chlorosome from the stressed culture, we varied the structural parameters of the aggregates: the lattice constants $a$ and $b$; the angle $\theta$ between the lattice vectors; and the orientation of the transition dipole. We also modified the shape of the aggregates by changing the length and the radius of the rolls, and also by transforming rolls to lamellae.

The absorption and CD spectra of the aggregates were computed using conventional relations (47), where the aggregate Hamiltonian was constructed within the point dipole approximation. For the value of the monomer transition dipole we used $|\mu|^2 = 30$ Debye (45). The calculated stick spectra were broadened using a Lorentz line with a half width of 100 $cm^{-1}$.

We find that for both structures I and II variations of the lattice constants within 3 – 5% or reorientations of the monomer transition dipoles within 3 – 5 degrees of the original values can result in a shift of the absorption spectra of the order of 100 $cm^{-1}$ which is comparable to the measured value. However, in order to change the type of the CD spectra the structural modifications should be more drastic. This result remains valid for all rolls we computed with the radius in the range $R$ = 3 – 11 nm and the length in the range $L$ = 10 – 130 nm.

The CD spectra of rolls with a large aspect ratio $L/R$ are of the mixed-type (–/+/–) for both structures I and II. Reducing the aspect ratio for the structure II we observe a transition between a mixed-type and type II (–/+) CD spectra, which agrees with the previous study (45). **Fig. 8*A*** shows the scaling of the transition length with the radius of a roll. Two types of the spectra for rolls of the same radius and different lengths are shown in **Fig. 8*B*.** For the structure I we obtain a transition to type I (+/–) for the roll length about 10 nm independently of its radius, **Fig. 9**. The latter type of CD spectra was reported for single mutant chlorosomes (48). It should be noticed that for some aspect ratios the rolls with the same values of $L$ and $R$ but different structures (I or II) give different type of CD spectra. For example, a roll with $R$ = 9 nm, $L$ = 80 nm and structure II has a type II CD spectrum, while the roll of the same shape but structure I has a mixed type of the spectrum. Thus, different types of CD spectra can also be explained by different folding of BChl aggregates.

Transitions between different types of CD spectra are also observed if we allow for the formation of curved lamella structures. In **Fig. 10**, we compare the CD spectrum of a roll and a curved lamella. The radius of the roll is $R = 7$ nm, and the length is $L = 100$ nm. The lamella is represented by a half of the roll with the same parameters cut along the symmetry axis. The spectra of both structures are calculated using the same formula (Eq. 13 in the report by Somsen et al. (1996) *Biophys J*, **71**, 1934-51) (47) and normalized to the number of molecules. For both types of BChl aggregation we obtained a transition between the mixed type and type II spectra. Within our model, **Fig. 6**, the contribution of lamellae and roll structures in CD spectra of the chlorosome should be controlled by the $L_2 / L_3$ ratio. While our SANS measurements do not provide information on how this ratio varies in the chlorosome from different cultures, changes in the cross-sectional radius of the chlorosome support that possibility.

To summarize the simulation results, we conclude that the shift of the $Q_y$ absorption maximum can be assigned to small variations in BChl aggregation. The changes in CD spectra should be associated with the modifications of the shape of the aggregate. In addition to a variation of the roll length proposed previously (45) as a possible cause for the change of CD type, we suggest that the transition can be due to formation of lamellae or different folding of aggregates (structure I vs. structure II). We notice that the structural modifications resulting in the change of the CD type cannot describe the blue-shift of the absorption peak.

*Electronic Excitation Spectra of Single BChl Homologues*
In addition to the structural analysis for the spectra described previously, we verified computationally that the changes in the frequency of the $Q_y$ electronic excitation of single BChls associated with the methylation at C-8 and C-12 positions cannot account for the blue shift of the absorption peak observed in the experiments. The calculations were done using time-dependent density functional theory as implemented in the Turbomole quantum chemistry package (49). We utilized the PBE0 (50) hybrid functional with the triple-ζ basis sets def2-TZVP (51). In order to simplify the calculations the farnesyl chain was substituted with a methyl group in all homologues.

The computed frequencies of BChl *c* $Q_y$ transitions, shown in **Fig. 11**, are blue shifted by about 2500 $cm^{-1}$ (approx. 15% of the excitation frequency) compared to the spectra of BChl *c* measured in methanol. This difference can be associated with the solvent effect and also to the systematic error of DFT calculations (52). The computed electron excitation frequency varies by about 100 $cm^{-1}$ for different homologues. Weighted by the homologues composition obtained from mass spectral analyses, **Fig. 1*B***, these can account only for about 10% (10 $cm^{-1}$) of the blue shift of the culture 6 absorption spectrum as compared to the culture 4.

**DISCUSSION**
Photosynthetic organisms use the energy and reducing equivalents generated via photosynthetic electron transport to produce building blocks of cells (i.e., via carbon metabolism) and undergo other cellular metabolic processes (e.g., nitrogen metabolism). It is known that carbon assimilation is a bottleneck of photosynthesis, and that enhancement of carbon assimilation is essential to the improvement of photosynthesis (15, 53, 54). The presented work shows that changing temperatures and carbon sources together can regulate the biosynthesis of BChl *c* homologues and the self-assembly of the chlorosome.

In this report we studied *Cba. tepidum* grown with various carbon sources at different temperatures and noted that culture 6, which is a temperature down-shifted culture grown on pyruvate/$HCO_3^-$, has some unique properties compared to the other cultures. These are:

(a) Combination of carbon assimilation and temperature shift exhibits more effects on chlorosome biogenesis (i.e., culture 6) compared with changing only one environmental factor.

(b) Slower growth of culture 6 may be due to more energy consumption in metabolic reactions.

(c) As culture 6 requires more energy and reducing equivalents produced from light-induced electron transport to produce acetyl-CoA and biomass via the reductive TCA cycle, less resource is available to synthesize BChl *c* homologues, which require significant energy and carbon sources (55, 56). Thus less BChl *c* homologues are incorporated in the chlorosome from culture 6.

(d) Culture 6 incorporates most BChl *c* homologues with a smaller substituent (i.e., an ethyl group) at C-8 attached to the chlorin ring in the chlorosome and produces a chlorosome with blue-shifted $Q_y$ absorption maxima. The composition of C-8 substituent has been suggested to contribute to the $Q_y$ transition band (34). Further, *bchQ* and *bchR* encode C-8- and C-12 methyltransferase, respectively, and *bchQ*(-), *bchR*(-) and *bchQ*(-)*bchR*(-) *Cba. tepidum* mutants were reported to exhibit a blue-shifted $Q_y$ maxima (57). We suggest that the blue-shift can be mainly assigned to small variations in the structure of BChl aggregates composing the chlorosome.

(e) Culture 6 producing most of BChl *c* homologues with ethyl group at C-8 and C-12 in the chlorosome suggests that C-12 methyltransferase is expressed, as C-12 can be either a methyl or ethyl group, and that C-8 methyltransferase is down-regulated as the C-8 substituent, which can be an ethyl, propyl, isobutyl and neopentyl group, is largely an ethyl group.

(f) In addition to affecting methylation at C-8 and C-12 substituents, culture 6 incorporates almost exclusively farnesyl alcohol (C15:3) plus very small amount of geranylgeranyl alcohol (C20:4) as the esterifying alcohol, whereas culture 4 and other cultures have slightly more geranylgeranyl alcohol (data not shown). It can be also explained as culture 6 has less energy and reducing equivalents available to synthesize a longer esterifying alcohol (i.e. a geranylgeranyl alcohol group). However, all cultures investigated in this paper synthesize farnesyl alcohol as the major esterifying alcohol of BChl *c*, as reported in literature (10, 12, 13, 23), so the change of the esterifying alcohol of BChl *c* in culture 6 is not as significant as the changes of C-8 and C-12 substituents.

(g) The chlorosome from culture 6 exhibits a type II (–/+) CD spectrum, whereas the chlorosome from other cultures we investigated in this report displays a mixed-type (–/+/–) CD spectrum. Previous theoretical studies suggest that the variation of CD spectra may arise from different lengths of cylindrical BChl aggregates (45, 58). As alternative mechanisms we propose that the change in CD type can occur due to different folding of aggregates or formation of curved lamellae.

(h) Culture 6 synthesizes a chlorosome with a smaller cross-sectional radius, which may partially arise from less BChl *c* homologues being synthesized and incorporated in the chlorosome. The smaller cross-sectional radius combined with the difference in the CD spectra may further indicate that the roll vs. curved lamellae composition of the chlorosome from culture 6 is different from other cultures.

(i) The chlorosome from culture 6 exhibits some similar properties to the chlorosome from *Cfl. aurantiacus*, including a type II (–/+) CD spectra (7, 21, 30), $Q_y$ absorption maxima (~740 nm) and simpler BChl *c* homologues with smaller C-8 and C-12 substituents. It remains to be understood the functional advantage of culture 6, which grows under stressed conditions, producing a chlorosome with a blue-shifted $Q_y$ absorption maxima.

(j) The 77K fluorescence spectra indicate that despite the blue shift of the chlorosome absorption peak the energy transfer from the BChl *c* energy domain to BChl *a* energy domain in the stressed culture may be more efficient.

(k) Finally, we would like to emphasize that for ambient light conditions carbon assimilation could be more sensitive to the amount of reaction centers and BChl *a* biosynthesis than the light-harvesting antenna complexes. Though this question goes beyond the scope of this paper, we may provide a qualitative analysis based on our data reported herein whereas details will need to be further investigated. Previous studies by Blankenship (59), Bryant (60, 61) and their coworkers estimate the presence of approximately 150 – 200 FMO trimers and 25 – 40 reaction centers per chlorosome molecule. Assuming that compared with other cultures the surface concentration of reaction centers in culture 6 (the stressed culture) remains the same and the area of the baseplate scales is proportional to the surface area of the chlorosome, we estimate that culture 6 contains approximately 40% less reaction centers compared with culture 4, which should be in the range 15 – 25 reaction centers per chlorosome.

## CONCLUSION

The biogenesis of the chlorosome has been previously reported to be regulated by light intensity, temperature, and electron sources, but no attention has been given to the issue of carbon assimilation, which is the bottleneck of photosynthesis. The present work suggests that carbon metabolism together with temperature changes regulate the biogenesis of the *Cba. tepidum* chlorosome, resulting in substantial structural modification of the chlorosome at different length scales. These structural variations may arise from the amount and types of BChl *c* homologues being synthesized and incorporated in the chlorosome. Moreover, our molecular modeling studies provide more detailed structural insight into the spatial arrangement of bacteriochlorophylls in the chlorosome.

**Acknowledgements**

The authors thank Dr. Xing Xu and Dr. Sun W. Tam at Nuclea Biotechnologies Inc. for assisting mass spectral measurements and Professor Gang Han at the University of Massachusetts Medical School for the access of his dynamic light scattering instrument. Bio-SANS CG-3 is a resource of the Center for Structural Molecular Biology at Oak Ridge National Laboratory supported by the U.S. Department of Energy, Office of Science, Office of Biological and Environmental Research Project ERKP291. Work in the laboratory of H.A.F. was supported by grants from the National Science Foundation (MCB-1243565) and the University of Connecticut Research Foundation. A.A.-G. and J.H. acknowledge support from the Center for Excitonics, an Energy Frontier Research Center funded by the US Department of Energy, Office of Science and Office of Basic Energy Sciences under award DESC0001088. A.A.-G. and S.K.S. acknowledge Defense Threat Reduction Agency grant HDTRA1-10-1-0046 and thank the Corning foundation for their generous support. J.K.T. is supported by start-up funds and faculty development fund from Clark University.

## Figure legends

**Figure 1. HPLC chromatogram and percentage of BChl *c* homologues in the chlorosome.** HPLC chromatogram of BChl *c* homologues in the chlorosome from cultures 1, 3, 4, 6, 7 and 9 with detection wavelength at 640 nm and the elution pattern of each BChl *c* homologue (A). Percentage of BChl *c* homologues in the chlorosome determined by mass peak area (B).

**Figure 2. SANS measurements for the chlorosome.** SANS profiles of the chlorosome isolated from constant temperature cultures at 50 $^{o}$C (culture 1 and culture 3), temperature down-shifted to 30 $^{o}$C cultures (culture 4 and culture 6) and constant temperature cultures at 30 $^{o}$C (culture 7 and culture 9) (A). SANS data fitted with modified Guinier's model (B) and with bi-axial ellipsoid of rotation (C).

**Figure 3. UV-visible spectra of cultures and the chlorosome**. Spectra of temperature down-shifted cultures grown on acetate/$HCO_3^-$ (culture 4) and pyruvate/$HCO_3^-$ (culture 6) (A); the chlorosome from culture 4 and culture 6 (B); and the chlorosome from culture 6 and *Cfl. aurantaicus* (C).

**Figure 4**. **CD spectra of the chlorosome.** CD spectra of the chlorosome from culture 4 and the chlorosome from culture 6 (A), the chlorosome from culture 6 and the chlorosome from *Cfl. aurantiacus* (B). The chlorosome with an OD of 1.0 for the BChl *c* $Q_y$ transition was used for the measurements.

**Figure 5**. **77 K fluorescence spectra**. 77 K fluorescence spectra of the chlorosome from culture 4 and culture 6 (A) and the whole cells from the same cultures (B). Spectra were normalized at the maxima of the chlorosome.

**Figure 6**. **Schematic illustration of BChl packing in the chlorosome.** The chlorosome is considered as an ellipsoidal body with semi-axes $L_{1-3}$. Two-dimensional aggregates of BChls are packed in a form of concentric rolls and curved lamellae.

**Figure 7. Model of BChl aggregation.** BChl pigments form two-dimensional lattice (Ganapathy et al. 2009, see ref. 35) with the lattice parameters $a = 6.25$ Å, $b = 9.79$ Å, $\theta = 122^{o}$, and $\alpha = 35^{o}$ (A). Binding of BChls is schematically illustrated. Brown arrows show the orientation of molecular $Q_y$ transition dipoles. (B) and (C) represent two suggested ways for aggregate folding in rolls. The structure I suggested for a mutant chlorosome with BChl stacks forming concentric rings (B) (ref. 41). The structure II proposed for a wild-type of bacteria (ref. 41), where BChl stacks are parallel to the roll symmetry axis of the roll (C).

**Figure 8. Structure II: length/radius dependence of CD spectra.** The characteristic length of a roll that corresponds to the transition between the mixed (–/+/–) and type II (–/+) CD spectra is shown as a function of a roll radius (A). Two types of CD spectra are shown for rolls of the radius $R = 7$ nm and the lengths $L = 100$ nm and $L = 60$ nm (B) marked by green and red stars in (A). The relative intensities are normalized to the number of molecules. The wavenumber scale shows the frequency red shift from the monomer transition due to BChl aggregation.

**Figure 9. Structure I: length dependence of CD spectra.** Two types of CD spectra are shown for rolls of the radius $R = 7$ nm and the lengths $L = 100$ nm and $L = 20$ nm. The transition

between the mixed type and type I spectra occurs independently of the roll radius at about $L = 20\,\text{nm}$.

**Figure 10. CD spectra: roll to lamella transition.** CD spectra of single rolls (radius $R$ = 7 nm and length is $L$ = 100 nm) as compared to the spectra of curved lamellae of the same curvature radius and length. The curved lamellae are represented by halves of the rolls cut along the symmetry axis. The relative intensities are normalized to the number of molecules. Structure I shows transition between the mixed type and type II spectra (A). Structure II shows transition between the mixed type and type II spectra (B).

**Figure 11. Computed $Q_y$ transition frequencies of BChl homologues.** The transition frequencies are computed using time dependent density functional theory. The following homologues are shown: *m/z* = 793.5 (8-ethyl, 12-methyl), *m/z* = 807.5 (8-ethyl, 12-ethyl), *m/z* = 823.5 (8-propyl, 12-ethyl), *m/z* = 839.5 (8-isobutyl, 12-ethyl) and *m/z* = 853.5 (8-neopentyl, 12- ethyl).

Fig. 1

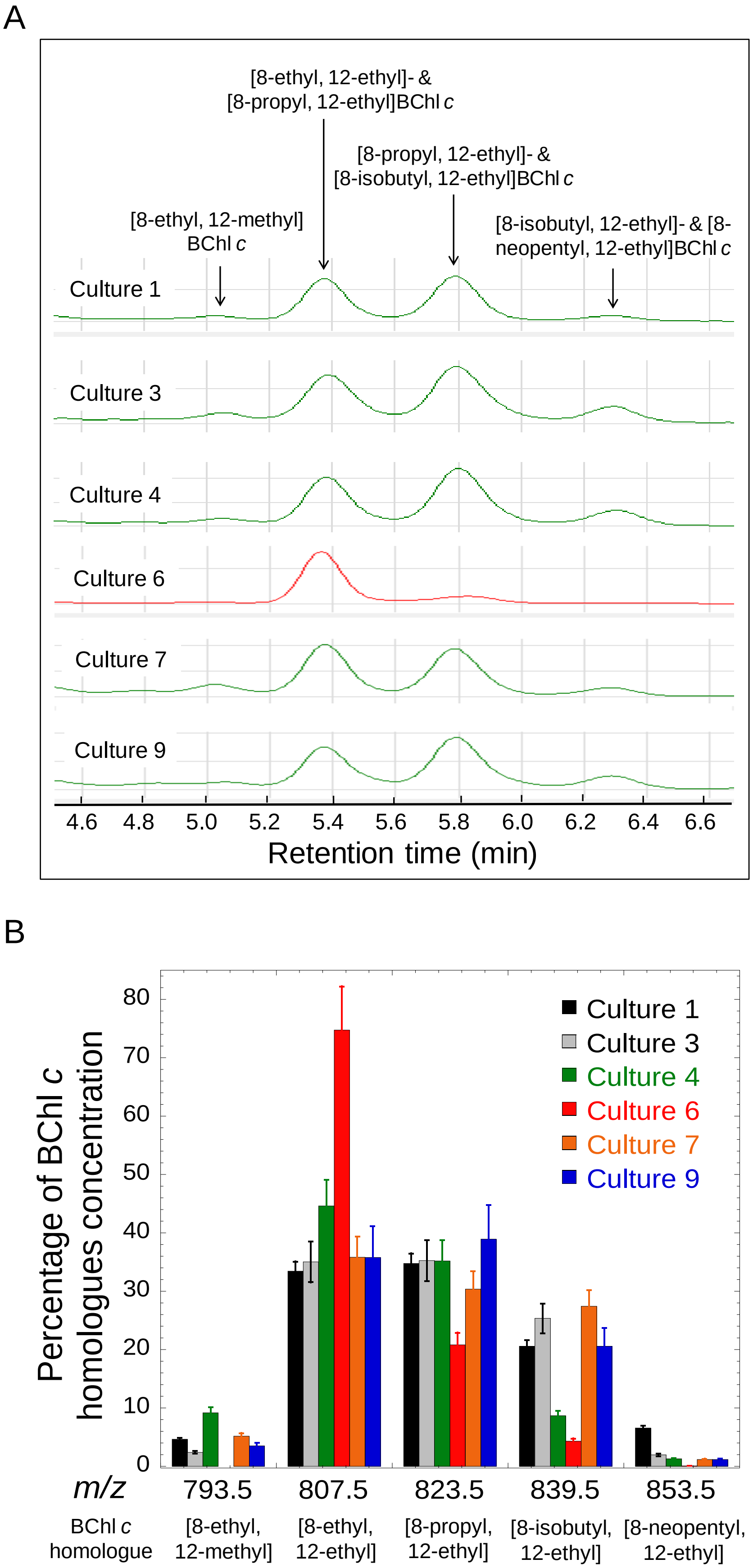

Fig. 2

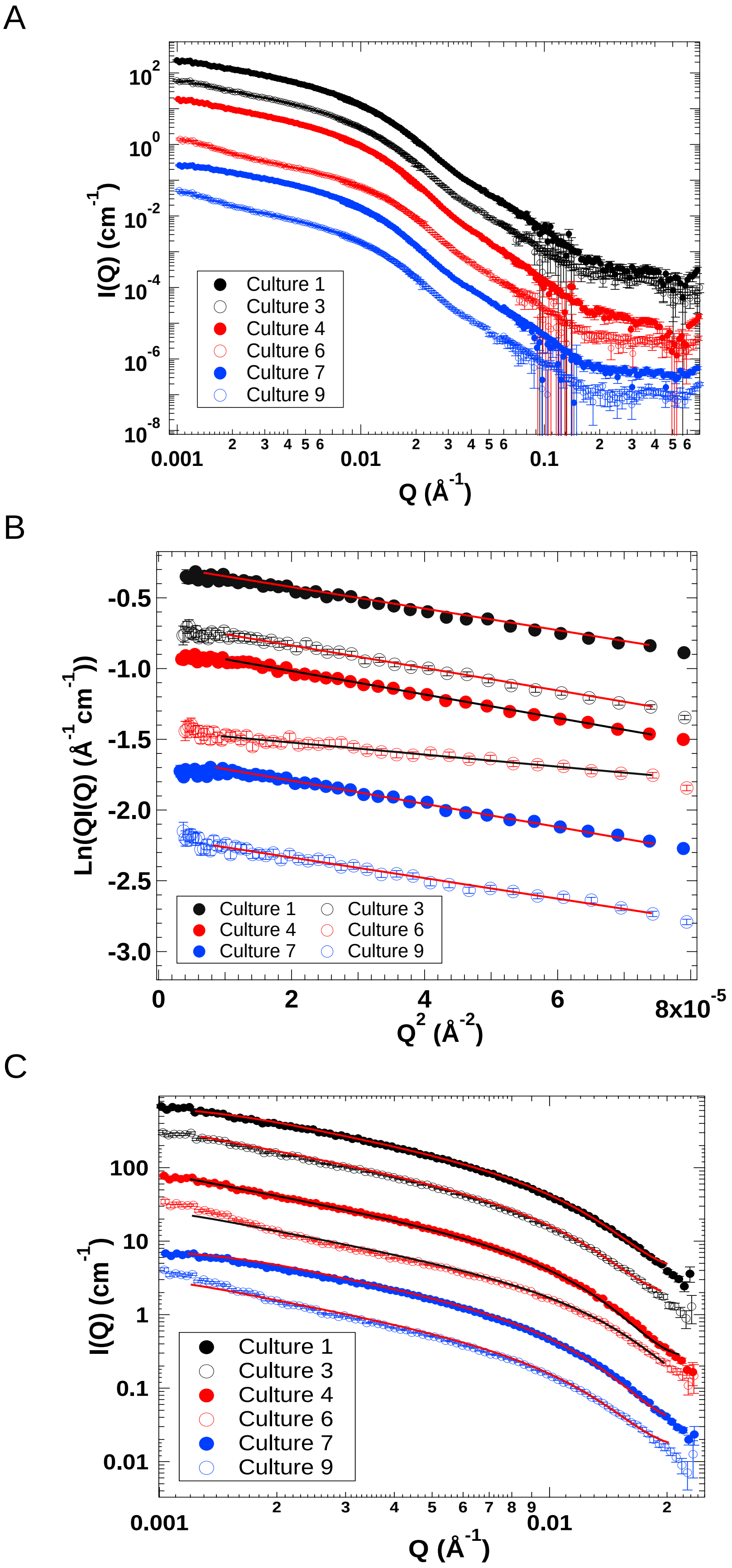

Fig. 3

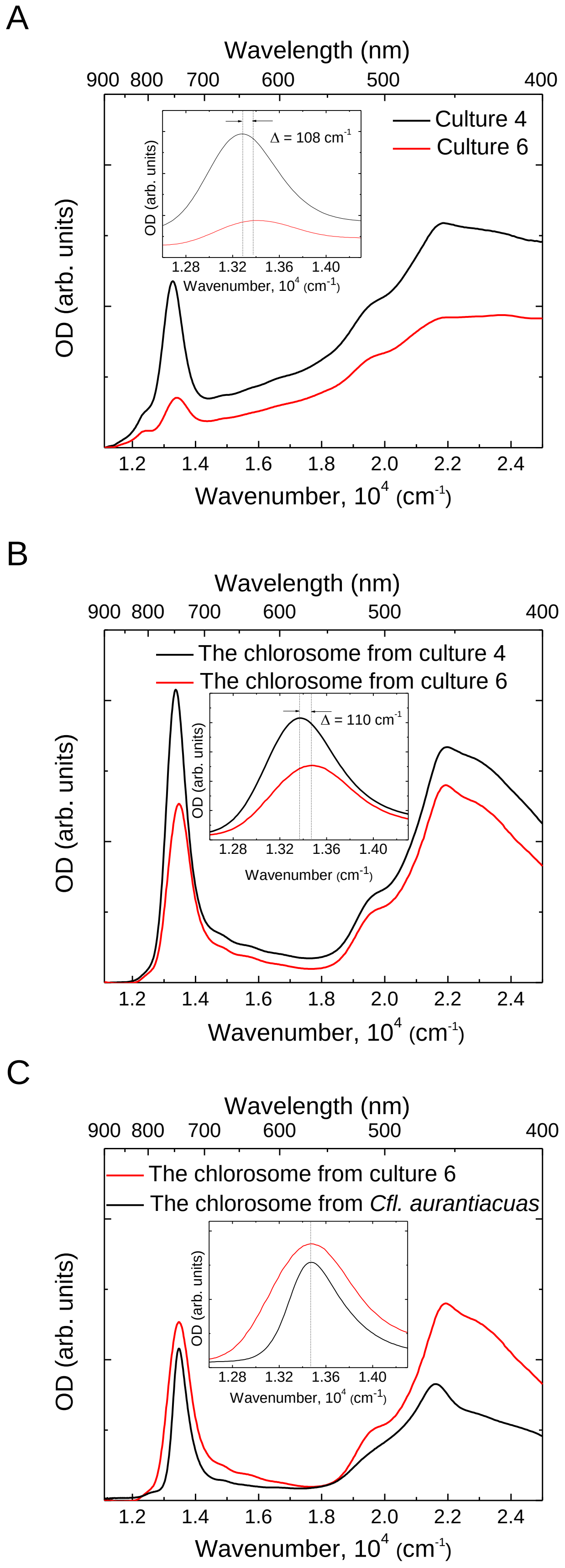
A
Wavelength (nm)
900 800 700 600 500 400
Culture 4
Culture 6
Δ = 108 cm-1
OD (arb. units)
1.28 1.32 1.36 1.40
Wavenumber, 10^4 (cm-1)
OD (arb. units)
1.2 1.4 1.6 1.8 2.0 2.2 2.4
Wavenumber, 10^4 (cm-1)
B
Wavelength (nm)
900 800 700 600 500 400
The chlorosome from culture 4
The chlorosome from culture 6
Δ = 110 cm-1
OD (arb. units)
1.28 1.32 1.36 1.40
Wavenumber (cm-1)
OD (arb. units)
1.2 1.4 1.6 1.8 2.0 2.2 2.4
Wavenumber, 10^4 (cm-1)
C
Wavelength (nm)
900 800 700 600 500 400
The chlorosome from culture 6
The chlorosome from Cfl. aurantiacuas
OD (arb. units)
1.28 1.32 1.36 1.40
Wavenumber, 10^4 (cm-1)
OD (arb. units)
1.2 1.4 1.6 1.8 2.0 2.2 2.4
Wavenumber, 10^4 (cm-1)

Fig. 4

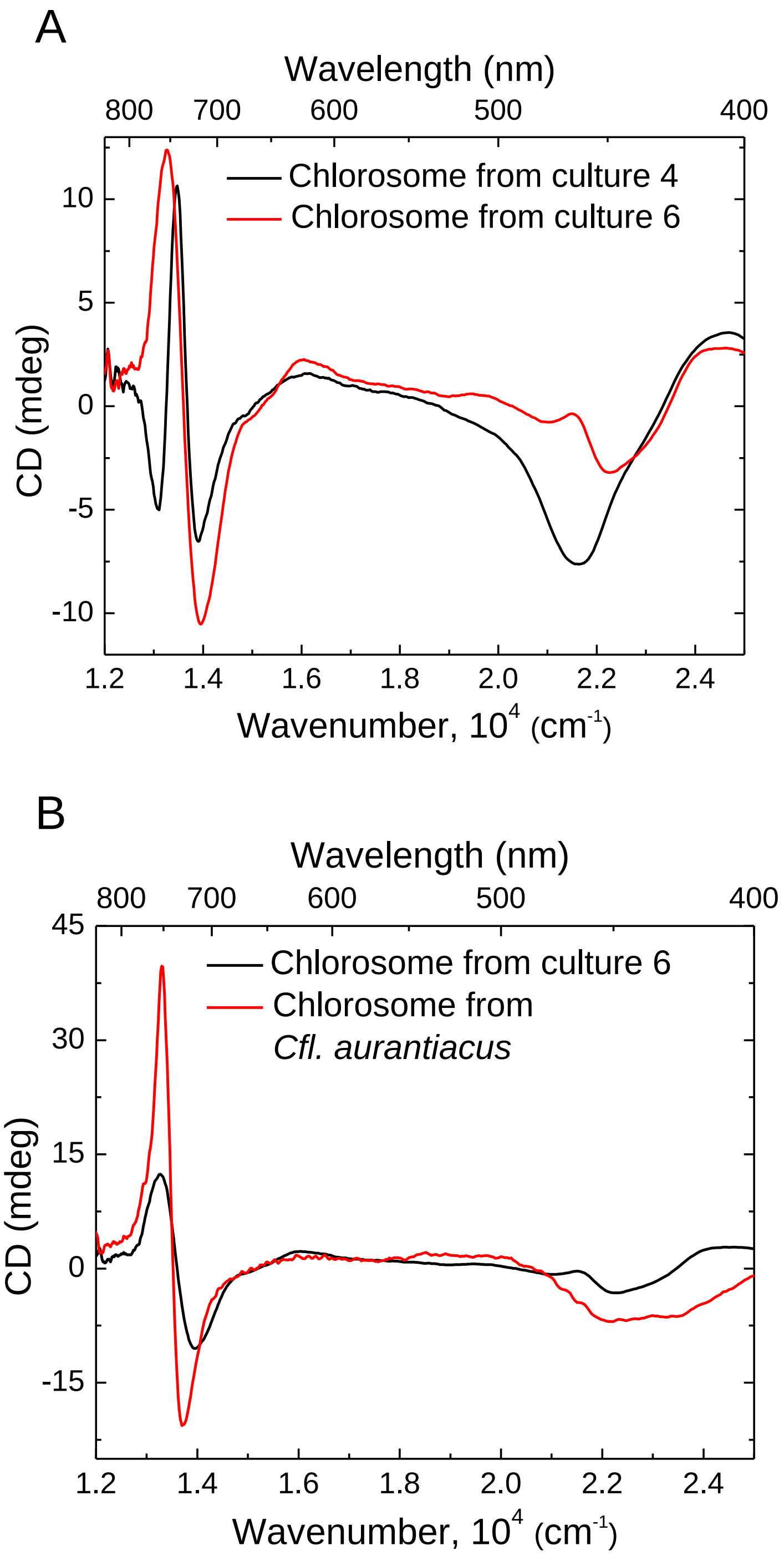

Fig. 5

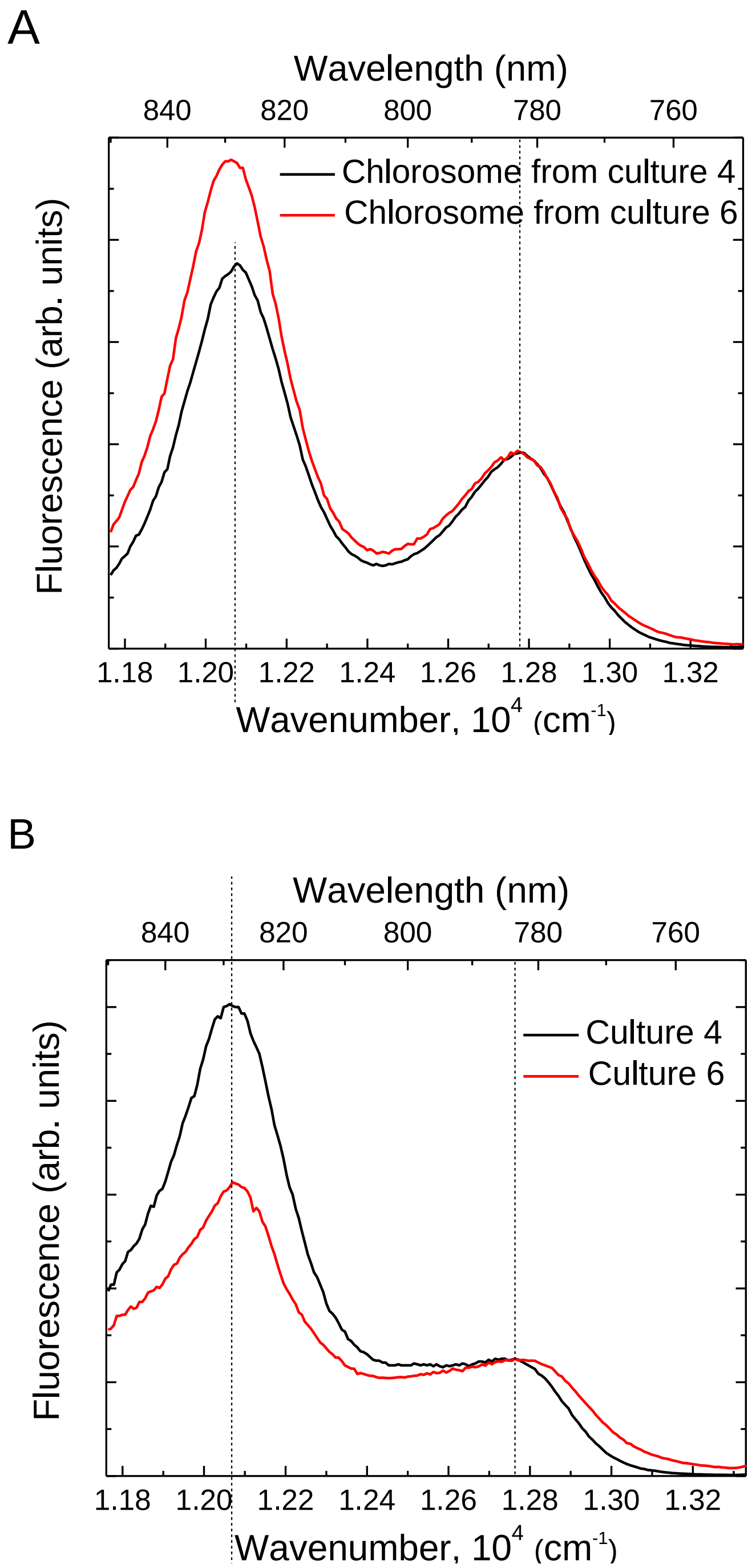

Fig. 6

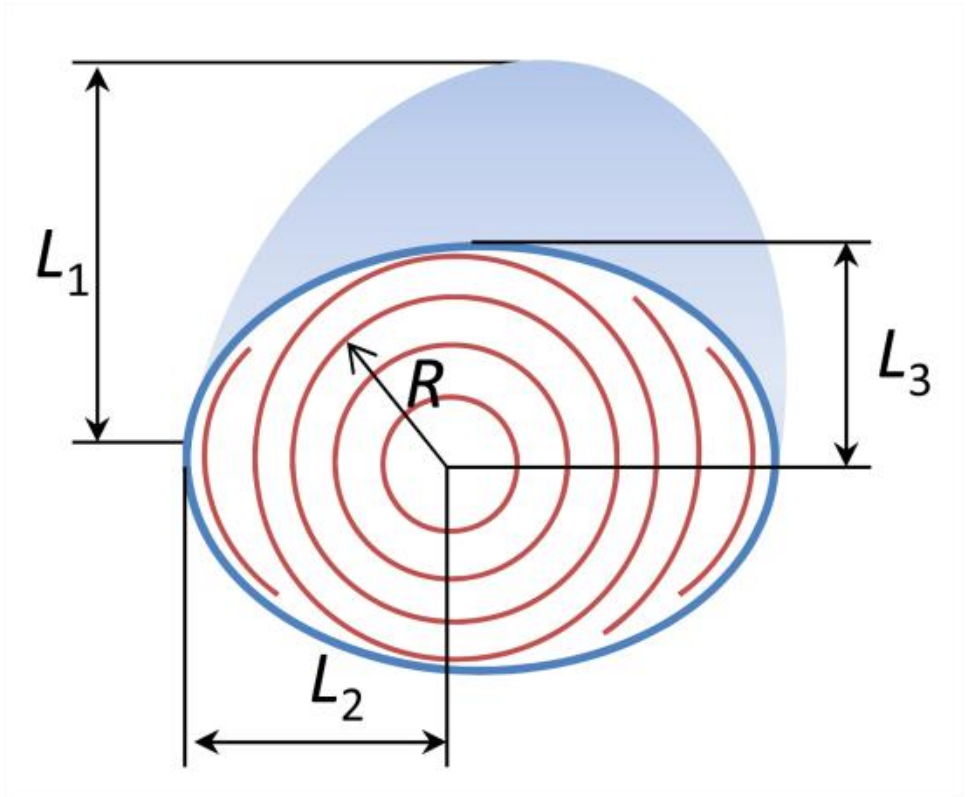

Fig. 7

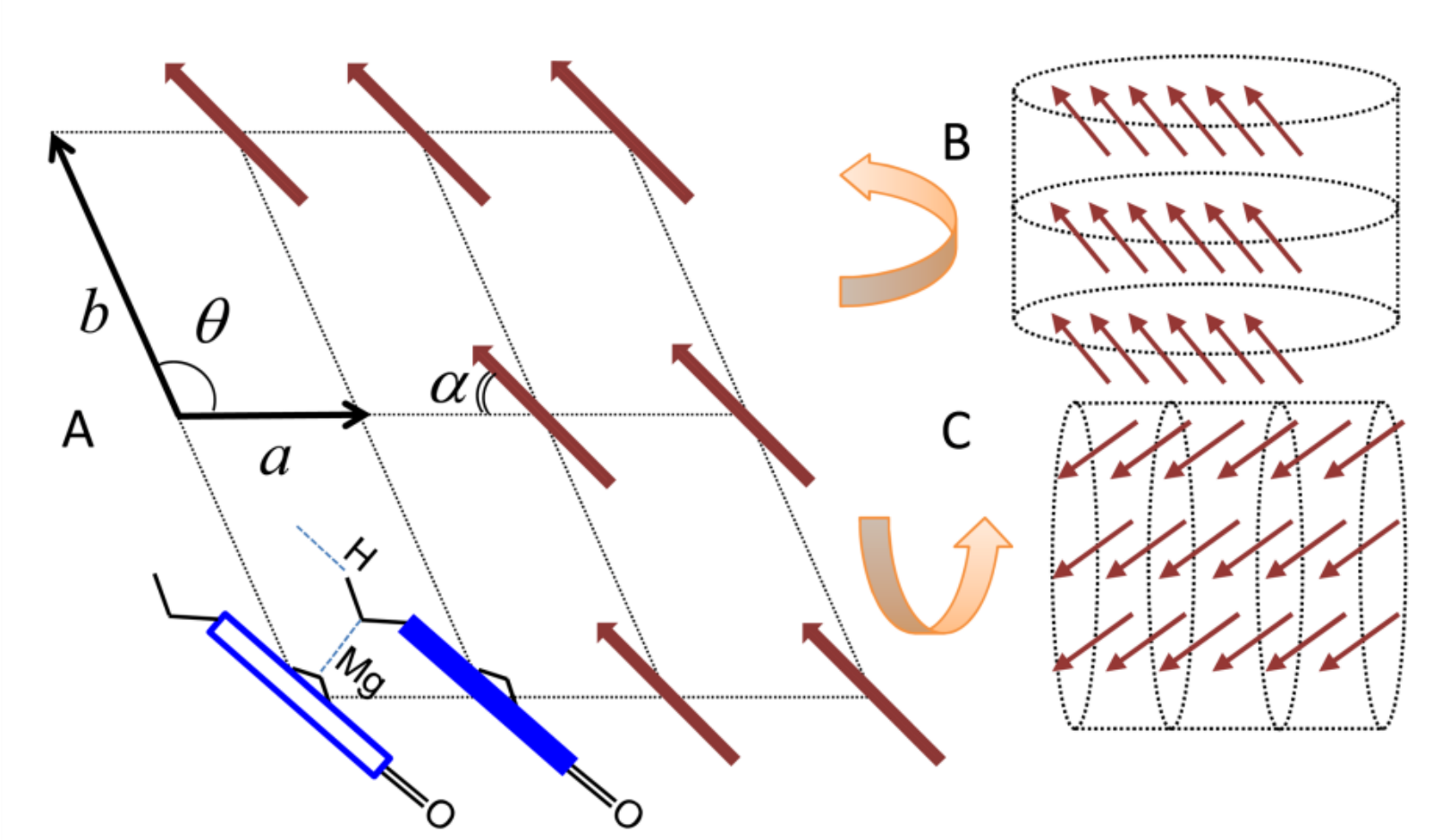

Fig. 8

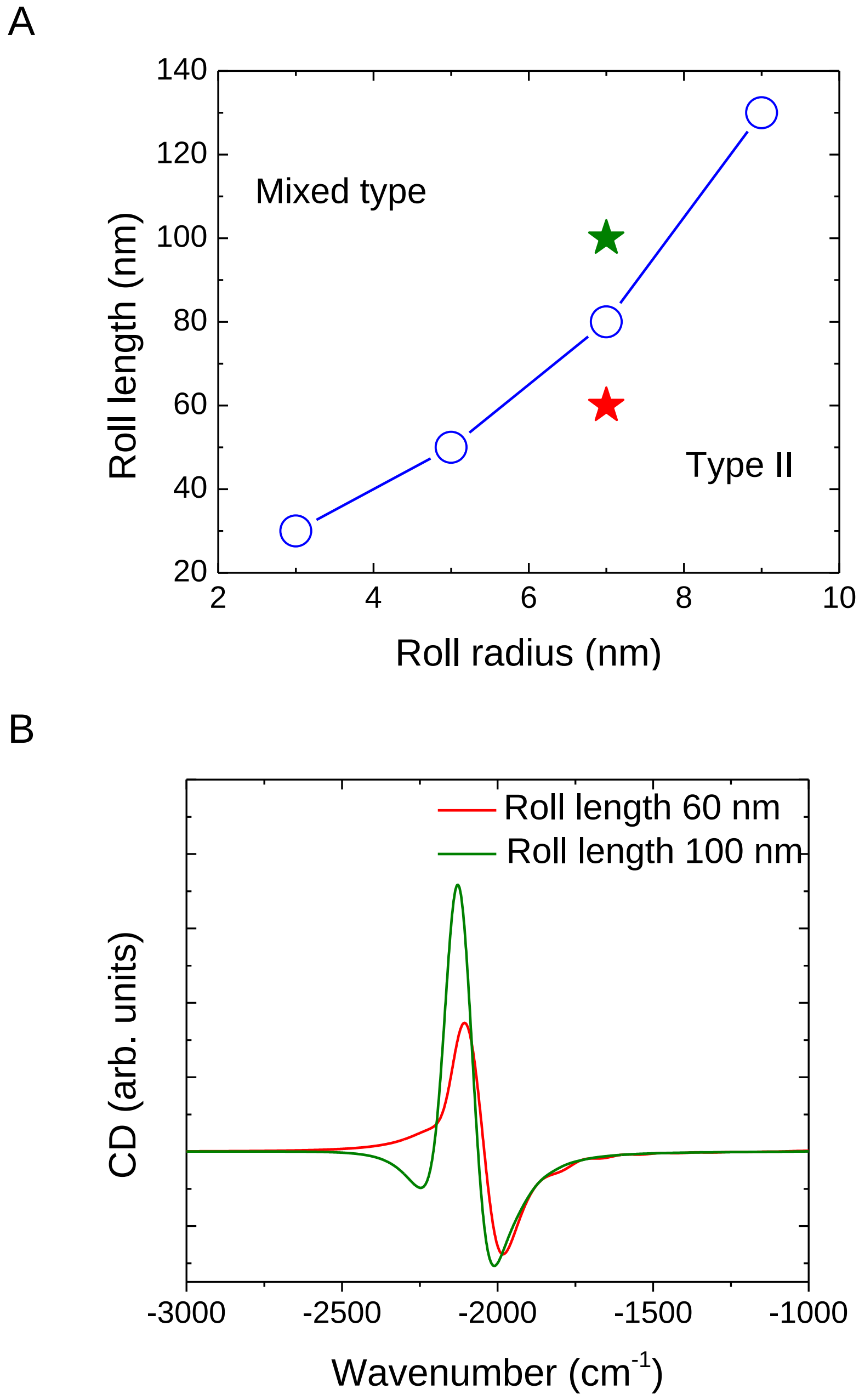

Fig. 9

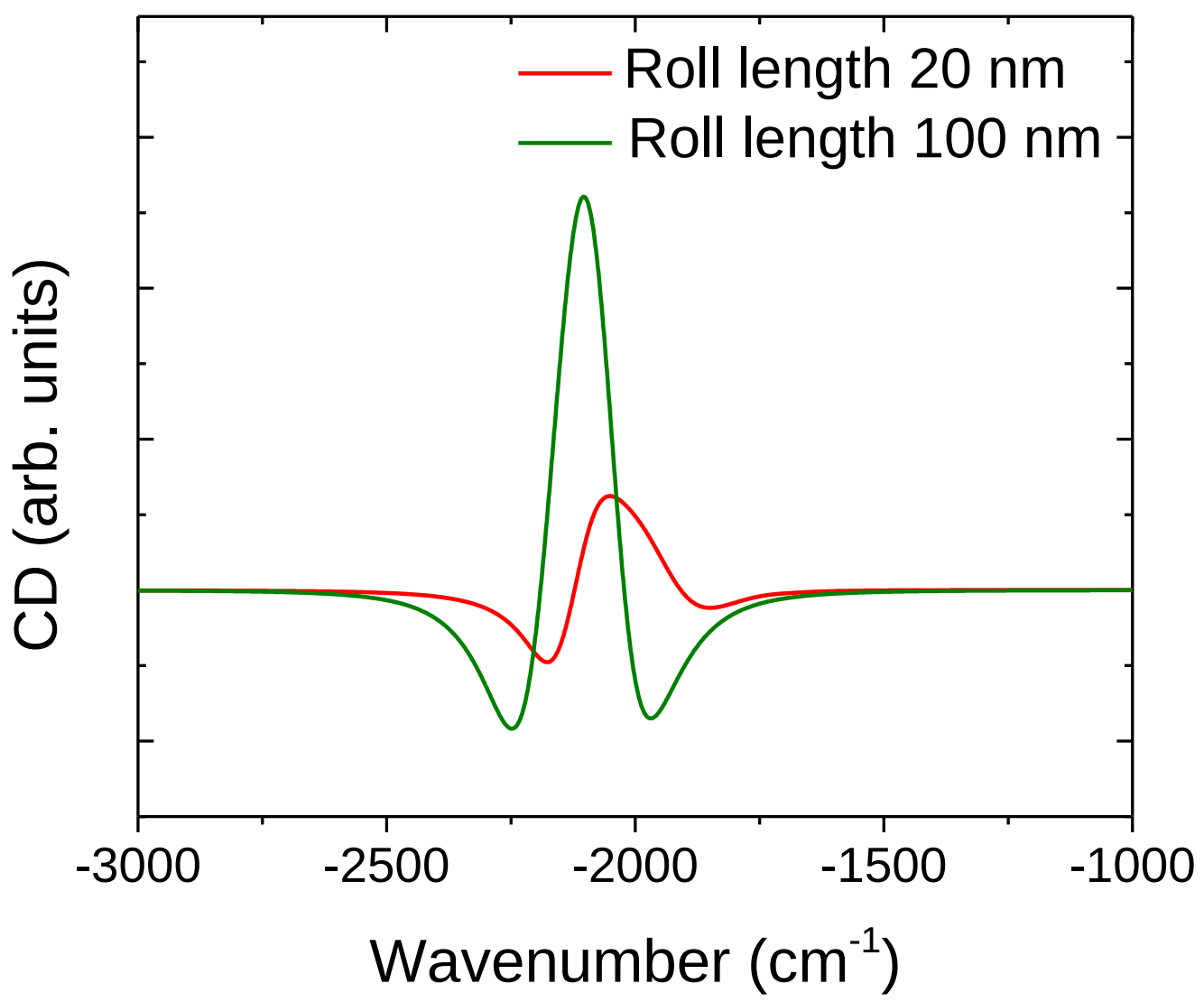

Roll length 20 nm
Roll length 100 nm
CD (arb. units)
-3000
-2500
-2000
-1500
-1000
Wavenumber (cm-1)

Fig. 10

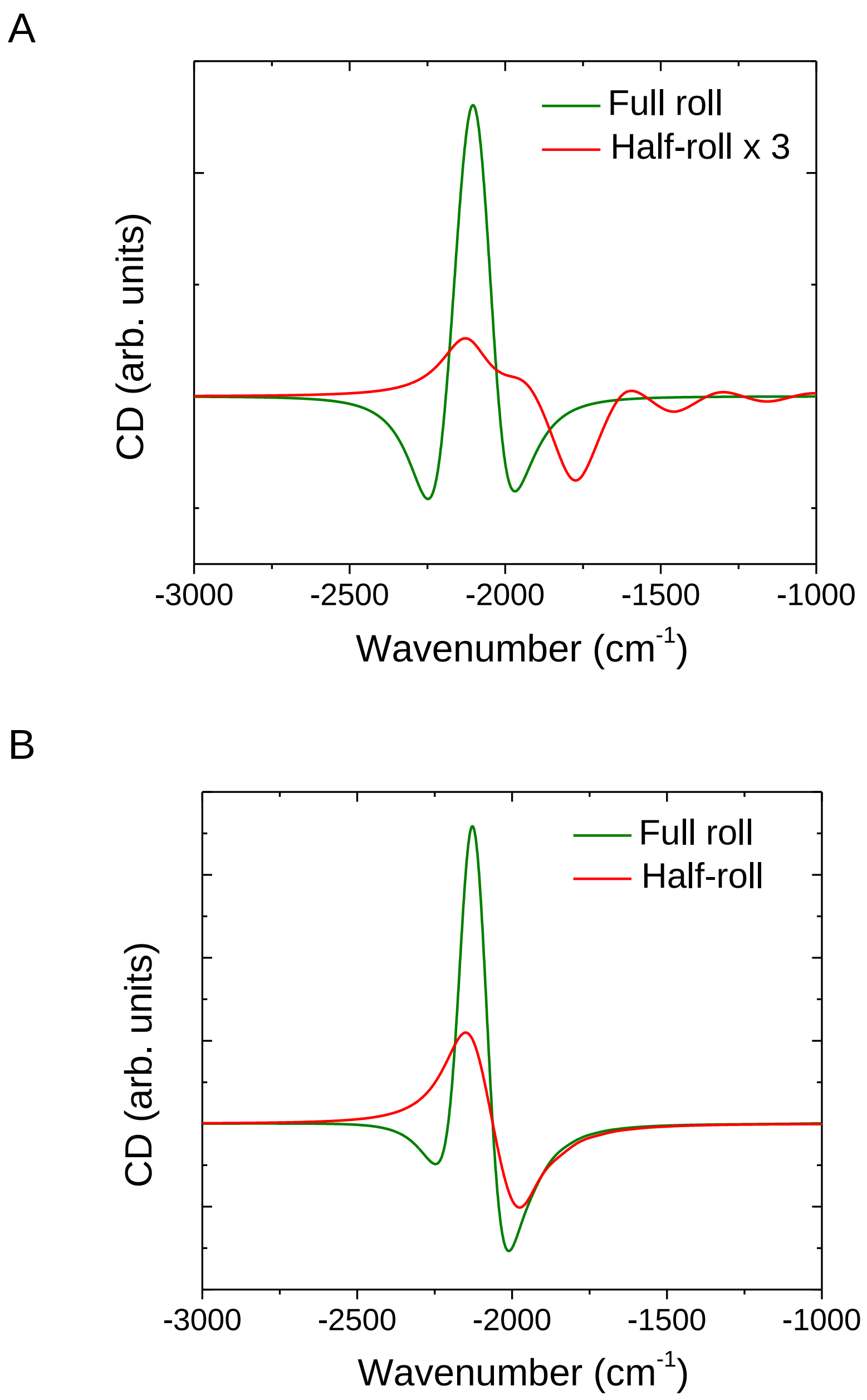

Fig. 11

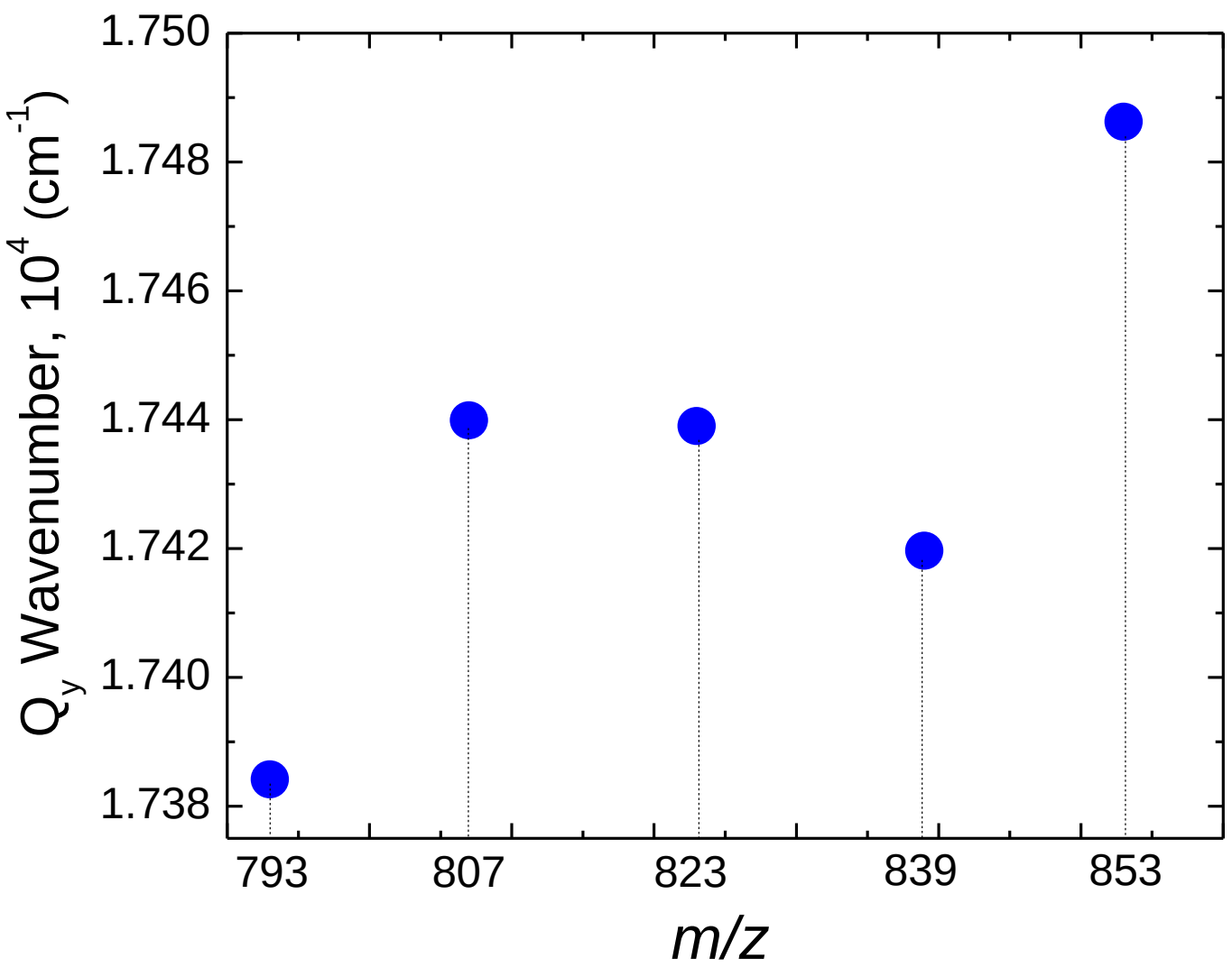

**Supporting Information**

# Temperature and Carbon Assimilation Regulate the Chlorosome Biogenesis in Green Sulfur Bacteria

Joseph Kuo-Hsiang Tang[1,]*, Semion K. Saikin[2], Sai Venkatesh Pingali[3], Miriam M. Enriquez[4], Joonsuk Huh[2], Harry A. Frank[4], Volker S. Urban[3], Alán Aspuru-Guzik[2]

[1]School of Chemistry and Biochemistry, Clark University, Worcester, MA 01610 USA, [2]Department of Chemistry and Chemical Biology, Harvard University, Cambridge, MA 02138 USA, [3]Center for Structural Molecular Biology, Biology and Soft Matter Division, Oak Ridge National Laboratory, Oak Ridge, TN 37831 USA, [4]Department of Chemistry, University of Connecticut, Storrs, CT 06269 USA

Running title: Metabolic regulation of chlorosome

*To whom correspondence should be addressed: Tel: 1-614-316-7886, Fax: 1-508-793-8861,

E-mail: jtang@clarku.edu

**Table S1. Cultural conditions reported in this paper**

| Cultures | Culture temperature | Carbon sources |
|---|---|---|
| Culture 1 | Constant temperature at 50 $^{o}$C | Acetate, bicarbonate |
| Culture 2 | Constant temperature at 50 $^{o}$C | Bicarbonate |
| Culture 3 | Constant temperature at 50 $^{o}$C | Pyruvate, bicarbonate |
| Culture 4 | Temperature down-shifted from 50 $^{o}$C to 30 $^{o}$C | Acetate, bicarbonate |
| Culture 5 | Temperature down-shifted from 50 $^{o}$C to 30 $^{o}$C | Bicarbonate |
| Culture 6 | Temperature down-shifted from 50 $^{o}$C to 30 $^{o}$C | Pyruvate, bicarbonate |
| Culture 7 | Constant temperature at 30 $^{o}$C | Acetate, bicarbonate |
| Culture 8 | Constant temperature at 30 $^{o}$C | Bicarbonate |
| Culture 9 | Constant temperature at 30 $^{o}$C | Pyruvate, bicarbonate |
| Culture 10 | Temperature up-shifted from 30 $^{o}$C to 50 $^{o}$C | Acetate, bicarbonate |
| Culture 11 | Temperature up-shifted from 30 $^{o}$C to 50 $^{o}$C | Bicarbonate |
| Culture 12 | Temperature up-shifted from 30 $^{o}$C to 50 $^{o}$C | Pyruvate, bicarbonate |

**Table S2. Values of growth rate, Qy absorption band, room temperature fluorescence peak and hydrodynamic diameter[1]**

| | | Thermal Stress → | | | |
|---|---|---|---|---|---|
| Carbon Source Stress ↓ | Temperatures / Carbon Sources | Constant temperature at 50 °C (50 → 50 °C) | Constant temperature at 30 °C (30 → 30 °C) | Temperature up-shifted to 50 °C (30 → 50 °C) | Temperature down-shifted to 30 °C (50 → 30 °C) |
| | Acetate + $HCO_3^-$ | Growth rate: 1.3 ± 0.1 ($h^{-1}$)<br>$Q_y$: 752 ± 1 nm<br>Fm: 767 ± 1 nm<br>$D_h$: 92 ± 5 nm | Growth rate: 1.0 ± 0.1 ($h^{-1}$)<br>$Q_y$: 751 ± 1 nm<br>Fm: 769 ± 1 nm<br>$D_h$: 94 ± 6 nm | Growth rate: 0.9 ± 0.1 ($h^{-1}$)<br>$Q_y$: 750 ± 1 nm<br>Fm: 768 ± 1 nm<br>$D_h$: 92 ± 3 nm | Growth rate: 0.7 ± 0.1 ($h^{-1}$)<br>$Q_y$: 751 ± 1 nm<br>Fm: 767 ± 1 nm<br>$D_h$: 92 ± 3 nm |
| | $HCO_3^-$ | Growth rate: 1.1 ± 0.1 ($h^{-1}$)<br>$Q_y$: 749 ± 1 nm<br>Fm: 764 ± 1 nm<br>$D_h$: 94 ± 3 nm | Growth rate: 0.8 ± 0.1 ($h^{-1}$)<br>$Q_y$: 749 ± 1 nm<br>Fm: 766 ± 1 nm<br>$D_h$: 92 ± 5 nm | Growth rate: 0.8 ± 0.1 ($h^{-1}$)<br>$Q_y$: 749 ± 1 nm<br>Fm: 766 ± 1 nm<br>$D_h$: 90 ± 3 nm | Growth rate: 0.6 ± 0.1 ($h^{-1}$)<br>$Q_y$: 748 ± 1 nm<br>Fm: 762 ± 1 nm<br>$D_h$: 88 ± 5 nm |
| | Pyruvate + $HCO_3^-$ | Growth rate: 0.8 ± 0.1 ($h^{-1}$)<br>$Q_y$: 748 ± 1 nm<br>Fm: 762 ± 1 nm<br>$D_h$: 88 ± 5 nm | Growth rate: 0.7 ± 0.1 ($h^{-1}$)<br>$Q_y$: 747 ± 1 nm<br>Fm: 764 ± 1 nm<br>$D_h$: 88 ± 4 nm | Growth rate: 0.7 ± 0.1 ($h^{-1}$)<br>$Q_y$: 747 ± 1 nm<br>Fm: 766 ± 1 nm<br>$D_h$: 86 ± 3 nm | Growth rate: 0.4 ± 0.1 ($h^{-1}$)<br>$Q_y$: 742 ± 1 nm<br>Fm: 756 ± 1 nm<br>$D_h$: 75 ± 4 nm |

[1]$D_h$, hydrodynamic diameter (measured by dynamic light scattering); Fm, room temperature fluorescence emission peak; $Q_y$, the maxima of the Qy absorption band

**Figure S1. Acetate and pyruvate assimilation in green sulfur bacteria.** Acetate is assimilated via the reductive TCA cycle and partial oxidative TCA cycle (A), and pyruvate is assimilated mainly via the reductive TCA cycle (B).

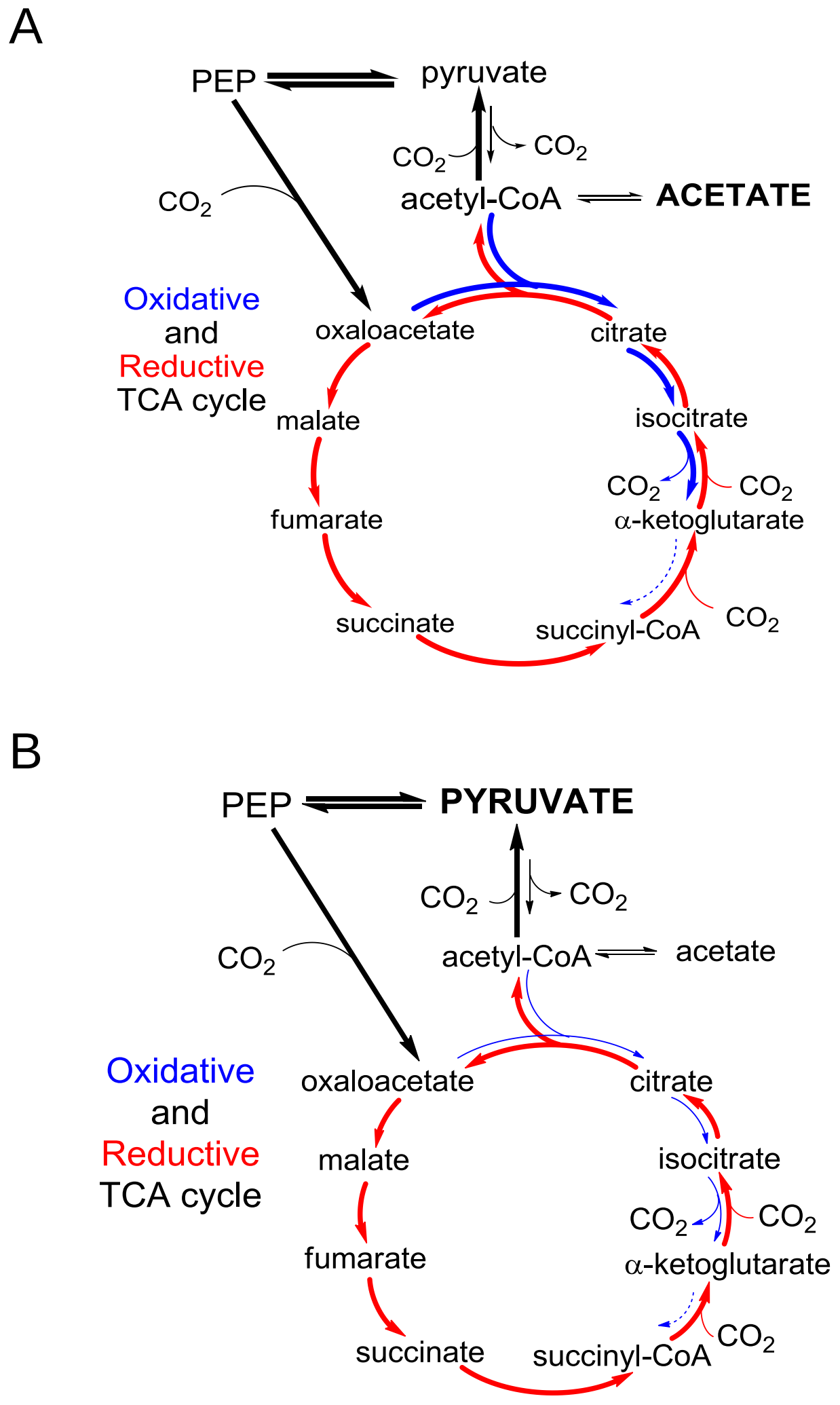

**Figure S2. Data fit for 77K fluorescence emission spectra shown in Fig. 6B.** Data fit were performed using Origin 8.6. Experimental data (solid line) and simulated data (dashed line) of chlorosome from culture 4 (A) and culture 6 (B) are shown.

A

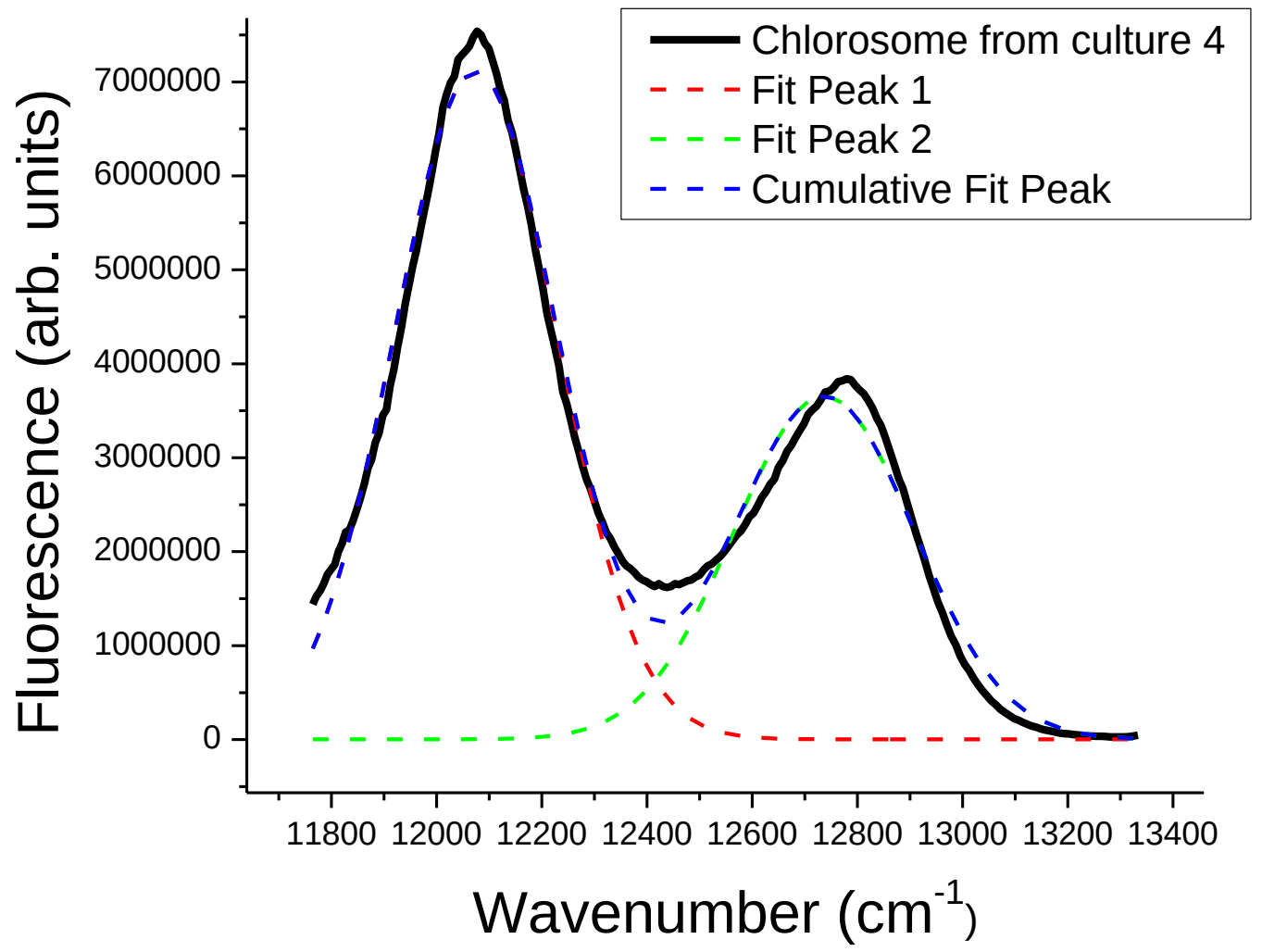

B

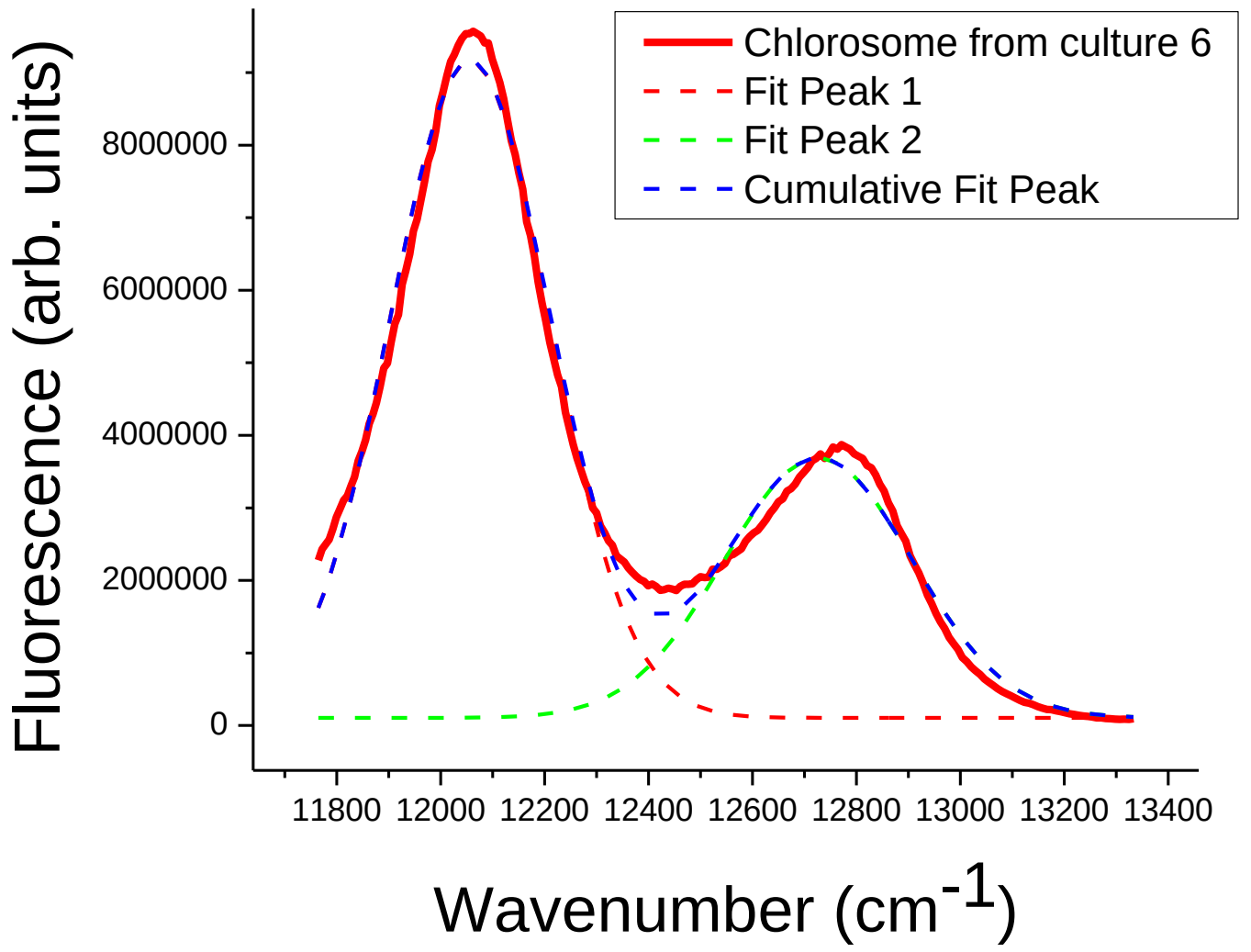

C

**Peak Area via Data Fit**

| Sample | Peak Area (chlorosome) | Peak Area (baseplate complex) | Ratio (= baseplate/chlorosome) |
|---|---|---|---|
| Chlorosome from culture 4 | 1.60E+09 | 2.80E+09 | 1.75 |
| Chlorosome from culture 6 | 1.60E+09 | 3.50E+09 | 2.19 |

**Figure S3. Data fit for 77K fluorescence emission spectra shown in Fig. 6B.** Data fit were performed using Origin 8.6. Experimental data (solid line) and simulated data (dashed line) of culture 4 (A) and culture 6 (B) are shown. Peak ratio is show in (C).

A

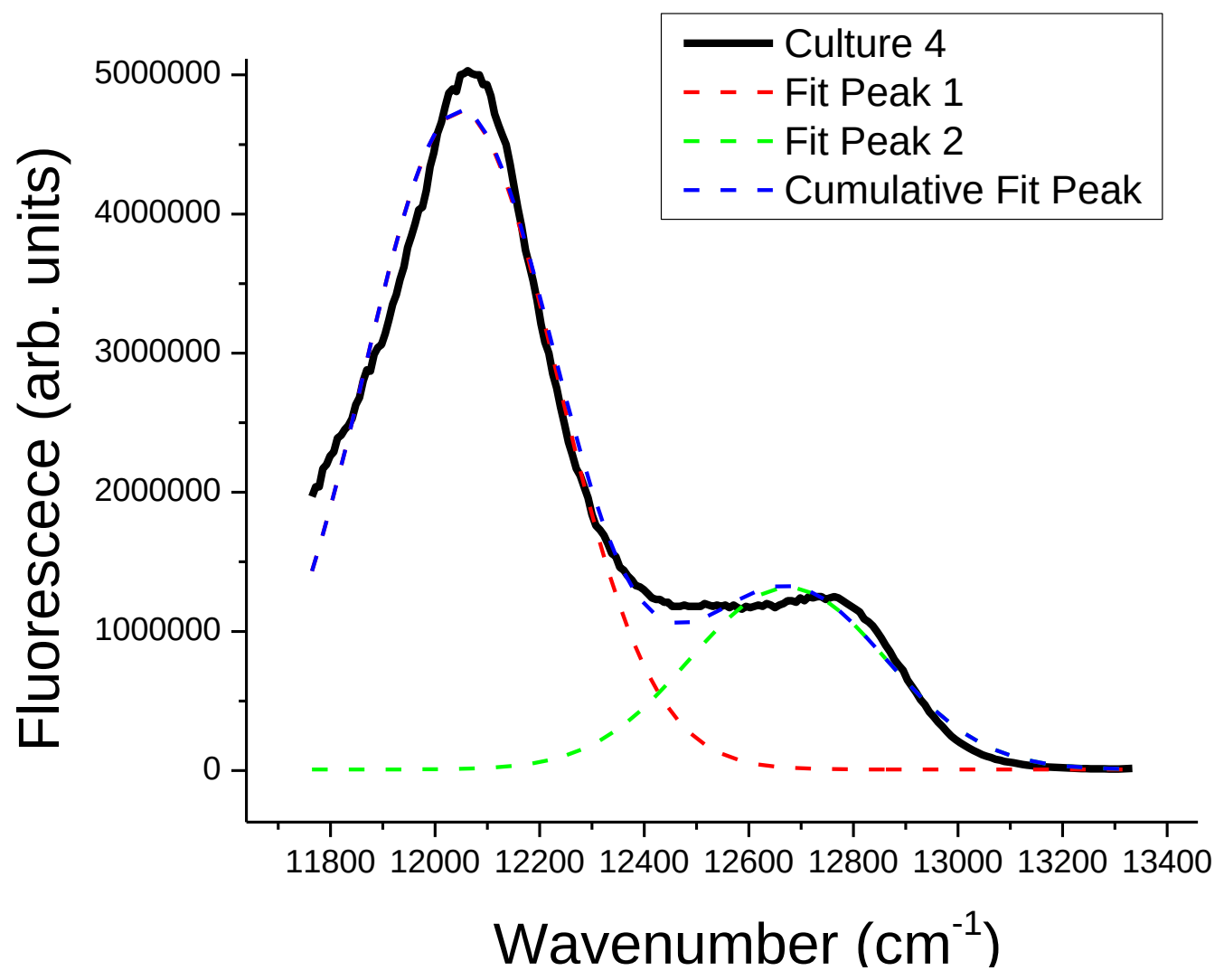

B

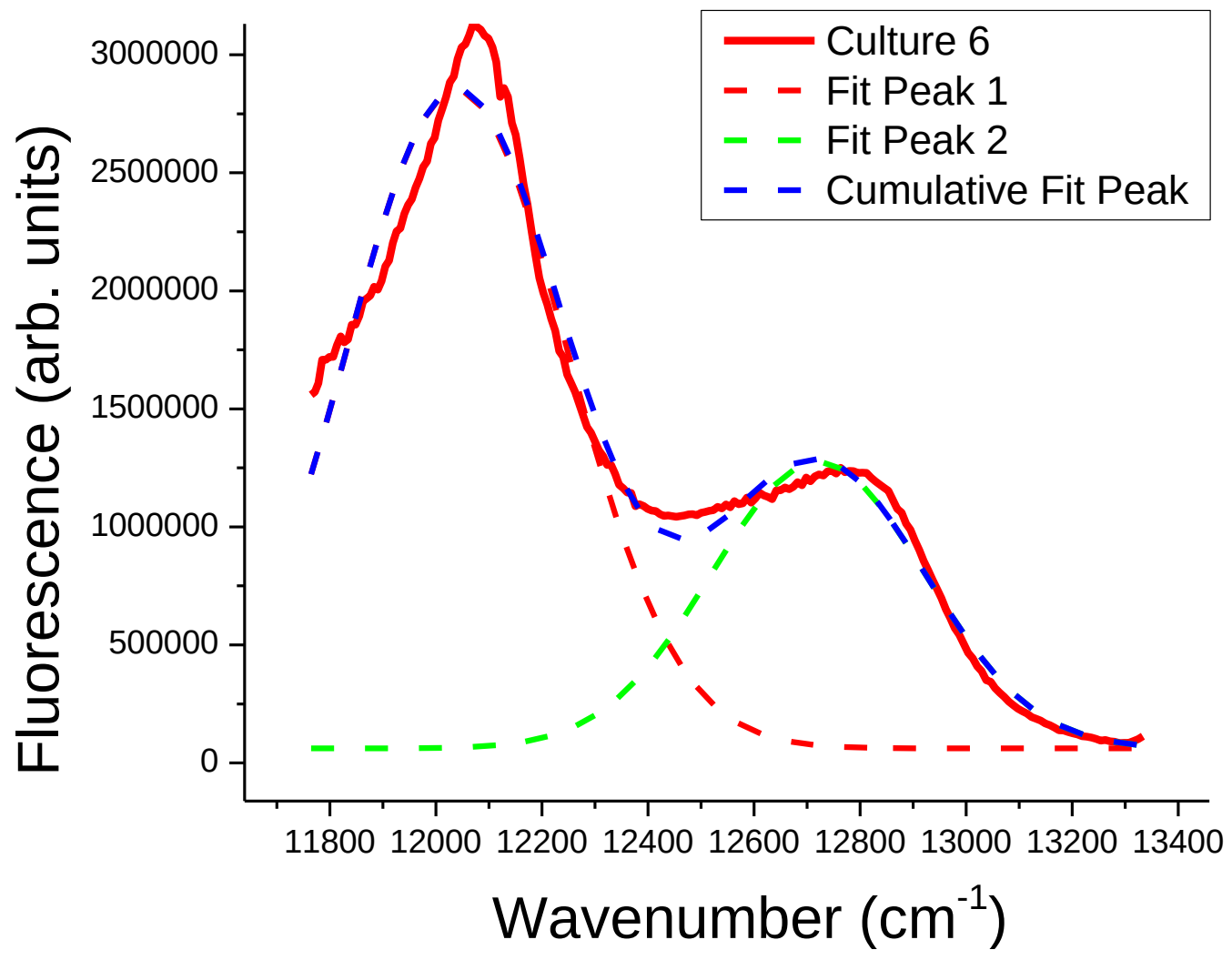

C

**Peak Area via Data Fit**

| Sample | Peak Area (chlorosome) | Peak Area (baseplate complex/the FMO protein) | Ratio (= chlorosome/baseplate and FMO) |
|---|---|---|---|
| **Culture 4** | 6.10E+08 | 2.20E+09 | 3.61 |
| **Culture 6** | 6.20E+08 | 1.45E+09 | 2.34 |